\documentclass[useAMS,usenatbib]{mn2e}

\usepackage{natbib}
\usepackage{amssymb}
\usepackage{longtable}
\usepackage[utf8]{inputenc}
\usepackage{supertabular}
\usepackage{graphicx}

\newcommand{\be}{\begin{equation}}
\newcommand{\en}{\end{equation}}

\def\zabs{$z_{\rm abs}$}

\def\mgii{Mg~{\sc ii}}

\def\ni2{Ni~{\sc ii}}
\def\nv{N~{\sc v}}

\def\aliii{Al~{\sc iii}}
\def\civ{C~{\sc iv}}
\def\siiv{Si~{\sc iv}}

\def\si2{Si~{\sc ii}}
\def\feii{Fe~{\sc ii}}
\def\h1{H~{\sc i}}
\def\feiii{Fe~{\sc iii}}

\def\kms{km s$^{-1}$}
\title[ \mgii\ BAL in SDSS J133356.02+001229.1]{Rapidly varying \mgii\ broad absorption line in SDSS J133356.02+001229.1}
\author[M. Vivek et al.]{M. Vivek$^{1,2}$\thanks{E-mail:vivekm@astro.utah.edu},R. Srianand$^{3}$ \& K.S. Dawson$^{2}$\\
$^{1}$ Dept. of Astronomy and Astrophysics, The Pennsylvania State University, 525 Davey Laboratory, University Park, PA 16802, USA\\
$^{2}$ Department of Physics and Astronomy, University of Utah, Salt Lake City, UT 84112, USA\\
$^{3}$Inter University Centre for Astronomy and Astrophysics, Pune 410007, India}
\begin{document}
\date{Accepted . Received ; in original form }
\pagerange{\pageref{firstpage}--\pageref{lastpage}} \pubyear{2002}
\maketitle

\label{firstpage}
\begin{abstract}
We report the discovery of rapid  variations of a high-velocity \mgii\ broad absorption line (BAL) trough in 
	the quasar SDSS J133356.02+001229.1 (z$_{em}$ $\sim$ 0.9197). \citet{vivek12} revealed the emergence and subsequent  near disappearance of a BAL component in this source having an ejection velocity of $\sim$ 28000 \kms.  Our further follow up studies with  South African Large Telescope (SALT) reveal the dramatic nature of the absorption line variability in this source. The absorption line emerged again at the same velocity and  nearly disappeared within the SALT observations. Our observations allow us to probe  variability   over  timescales   of the order of few days to 4.2 years in the QSO rest-frame. The observed velocity stability of BAL absorption does not point to any line of sight acceleration/deceleration of BAL clouds. The ionization parameter of the absorbing cloud is constrained from the column density ratio of \mgii\ to \feii\ ground state absorption. In the absence of strong optical continuum variability, we suggest that photoionization driven BAL variability due to changes in the shielding,   multiple streaming clouds across our line of sight in a co-rotating wind  or a combination of both as possible explanations for the observed strong equivalent width variations. 
\end{abstract}


\section{Introduction}
The radiative mode of active galactic nuclei (AGN) feed back, operated through outflowing winds,  has been proposed to be the most likely  explanation for the observed  supermassive black hole (SMBH)-host galaxy bulge co-evolution and the star formation process  \cite[e.g.,][]{dimatteo05,higginbottom13}. The most direct evidence of disk winds in AGNs is provided by broad absorption line quasars (BALQSOs). These objects exhibit blue-shifted broad absorption lines, at least 2000 \kms\ wide, associated  with  strong  resonance  lines  in  the  ultraviolet  (UV)  wavelengths. A vast majority of BALQSOs belong to the subclass called high ionization BAL (HiBAL) QSOs which only contain BALs of certain high ionization lines like \nv, \siiv\ and \civ.  About 15 per cent of BALQSOs also show low ionization lines like \mgii\ and \aliii\ together with the high ionization lines and are called low ionization BAL (LoBAL) QSOs. An even rarer population of BALQSOs ($\sim$ 1 per cent) also contains broad absorption from excited fine-structure levels of iron  which are known by FeLoBAL QSOs.    
\begin{table*}
\caption{Log of SALT observations:    }
\begin{tabular}{|c|c|c|c|c|c|c|c|c|c|}
\hline
 Name      &		Date	        &Grating&Cam-Angle&Exp.time&Airmass& Wavelength & Resolution & SNR$^{a}$	\\	
           &            (YYYY-MM-DD)    &	& (degree)&   (s) &       &   ($\AA$)  &            &      \\ 
\hline
			& 2012-05-09	&PG0900& 25.75&	1200&	1.283 &	3320-6440 & 944  &36.19	\\ 
			& 2012-05-09	&PG0900& 25.75&	1200&	1.254 &	3320-6440 & 944  &42.40 	\\ 
			& 2012-05-31	&PG0900& 25.75&	1200&	1.330 &	3320-6440 & 944  &51.79 	\\ 
			& 2012-05-31	&PG0900& 25.75&	1200&	1.399 &	3320-6440 & 944  &50.64 	\\ 
			& 2012-06-01	&PG0900& 26.5 & 1200&	1.530 &	3460-6576 & 911  &52.89 	\\ 
			& 2012-06-01	&PG0900& 26.5 & 1200&	1.270 &	3460-6576 & 911  &56.84 	\\ 
			& 2012-06-03	&PG0900& 25.75&	1200&	1.348 &	3320-6440 & 944  &54.84 	\\ 
			& 2012-06-03	&PG0900& 25.75&	1200&	1.205 &	3320-6440 & 944  &56.93 	\\ 
			& 2013-05-01	&PG0900& 25.75&	1200&	1.284 &	3320-6440 & 944  &54.64 	\\ 
			& 2013-06-23	&PG0900& 25.75&	1200&	1.293 &	3320-6440 & 944  &41.14 	\\ 
			& 2013-06-23	&PG0900& 25.75&	1200&	1.350 &	3320-6440 & 944  &41.58 	\\ 
			& 2014-02-17	&PG0900& 25.75&	1300&	1.269 &	3320-6440 & 944  & 9.56 	\\ 
			& 2014-02-17	&PG0900& 25.75&	1300&	1.230 &	3320-6440 & 944  & 9.46 	\\ 
			& 2014-02-27	&PG0900& 25.75&	1000&	1.334 &	3320-6440 & 944  &50.25 	\\ 
			& 2014-03-13	&PG0900& 25.75&	1200&	1.247 &	3320-6440 & 944  &45.01 	\\ 
J133356.02+001229.1	& 2014-03-14	&PG0900& 25.75&	1009&	1.188 &	3320-6440 & 944  &28.84 	\\ 
			& 2014-03-14	&PG0900& 25.75&	747 &   1.202 &	3320-6440 & 944  &33.16 	\\ 
			& 2014-04-11	&PG0900& 25.75&	1300&	1.234 &	3320-6440 & 944  &52.69 	\\ 
			& 2014-04-11	&PG0900& 25.75&	1300&	1.274 &	3320-6440 & 944  &54.84 	\\ 
			& 2014-05-21	&PG0900& 25.75&	1300&	1.221 &	3320-6440 & 944  &55.32 	\\ 
			& 2014-05-21	&PG0900& 25.75&	1300&	1.257 &	3320-6440 & 944  &52.01 	\\ 
			& 2014-06-21	&PG0900& 25.75&	900 &   1.251 &	3320-6440 & 944  &53.38 	\\ 
			& 2014-06-21	&PG0900& 25.75&	900 &   1.283 &	3320-6440 & 944  &57.99 	\\ 
			& 2014-06-21	&PG0900& 25.75&	900 &   1.322 &	3320-6440 & 944  &48.26 	\\ 
			& 2015-03-14	&PG1300& 37.75&	2100&	1.193 &	3900-5990 & 824  &53.35 	\\ 
			& 2015-04-20	&PG1300& 37.75&	1700&	1.235 &	3900-5990 & 824  &41.46 	\\ 
			& 2015-06-08	&PG0900& 25.75&	1700&	1.207 &	3320-6440 & 944  &64.05 	\\
			& 2016-02-11	&PG0900& 25.75&	1700&	1.207 &	3320-6440 & 944  &50.53 	\\
			& 2016-03-14	&PG0900& 25.75&	1700&	1.207 &	3320-6440 & 944  &56.94 	\\
			& 2016-04-13	&PG0900& 25.75&	1700&	1.207 &	3320-6440 & 944  &76.89 	\\

\hline                                                                                              		  
\hline                                                                                            
\end{tabular}                                                                                     
 \begin{flushleft}
$^{a}$ SNR per pixel estimated between the wavelength ranges 4500--4800~\AA. 
  \end{flushleft}
\label{obs_tab}
\end{table*}

Absorption line variability studies of BALs are an important tool for understanding the gas dynamics occurring close to the central engine. Most of the previous variability studies of BALs mainly concentrated on  high ionization lines like \civ\ and \siiv\ \citep{lundgren07,gibson08,capellupo11,filiz12,filiz13,vivek14,welling14}. This is partly due to the availability of a larger sample of HiBALs and partly due to the difficulty in disentangling the true BAL variability from Fe emission variability in  LoBALs.  \citet{vivek12a} probed the time variability of five FeLoBALs spanning an interval of up to 10 years in the quasar rest frame and found strong variations of fine-structure Fe II UV 34 and UV 48 lines in the spectra of SDSS J221511.93-004549.9. \citet{vivek14} reported that LoBALs are found to be less variable compared to HiBALs in their spectroscopic monitoring study using  27 LoBALs. Although  BALQSOs are known to vary in their absorption troughs, the most interesting cases of BAL variation are when the trough variability (1) exhibits signatures of radiative acceleration \citep[for e.g.,][]{anand02,grier16} or (2) is driven by photoionization \citep[for e.g.,][]{wang15}. Equally interesting are the cases where the BAL  completely disappears or appears between two observations. There has been a handful of individual studies reporting disappearance/appearance of \civ\ BAL transients \citep{ma02,hamann08,leighly09,Krongold10,hidalgo11}. \citet{filiz12} reported 19 cases of BAL trough disappearance in 21 sources in their studies using  Baryon Oscillation Spectroscopic Survey \citep[BOSS,][]{dawson13} data. \citet{vivek16} searched for transient BALs in a sample of 50 HiBALs and reported 6 cases of BAL appearance/disappearance. \citet{mcgraw17} found  14 disappearing BALs and 18 emerging BALs from their search of 470 quasars having multi-epoch observations in Sloan Digital Sky Survey (SDSS). All the previous studies attributed the BAL transience  either to  multiple streaming wind moving across the line of sight or to ionization-change scenario.

The first case of BAL transience in a LoBAL QSO was reported by \citet{vivek12} in  SDSS J133356.02+001229.1 (hereafter J1333+0012 at ${z_{em} = 0.9197}$). In this previous study, \citet{vivek12} identified two BAL components centered at  ejection velocities of 17,000~\kms\ (R component) and 28,000~\kms\ (B component) in the SDSS spectra obtained in 2001.  During our spectroscopic monitoring campaign using 2-m telescope at IUCAA Girawali Observatory (IGO)  between 2008 and 2011,  the R component completely disappeared  whereas the B component emerged, strengthened in optical depth, widened in velocity and {nearly disappeared in 2011}.
In this paper, based on our continuous monitoring of this source, we study  variability  exhibited by this \mgii\ BAL component over  timescales   of the order of few days to 4.2 years in the QSO rest-frame. 
From 2008 onwards,  this B component has shown dramatic variability between spectra obtained on any consecutive years. This article is organized as follows. In section 2, we provide the details of our spectroscopic observations and data reduction. This section also provides some details of Catalina Real-Time Transient Survey (CRTS) data used to quantify the continuum variability of this quasar.  In section 3, we provide the statistical analysis of BAL variability. In section 4, we discuss the observed variability in the frame work of different BALQSO models. Our results are summarized in section 5.

\section{Observation \& Data Reduction}
Our spectroscopic observations were primarily carried out with the IUCAA Faint Object Spectrograph (IFOSC) mounted on a 2-m telescope at IUCAA Girawali Observatory (IGO) and the Robert  Stobie  Spectrograph  \citep[RSS,][]{kobulnicky03,smith06}  on  South African Large Telescope (SALT). We also used two archival spectra available from the SDSS data release 12 \citep{alam15}. The details about the  IGO observations and data reduction are given in \citet{vivek12}. Briefly, we used IFOSC to obtain  high signal-to-noise spectra  every year from  2008 to 2011 with R$\sim$ 1000 and wavelength coverage 3200--6800~\AA. The two SDSS spectra obtained on 2001 and 2003 have R$\sim$ 2000 and  a wavelength coverage of 3800--9200~\AA.
 
We continued our monitoring campaign using RSS  on SALT from  2012 to  2016. We used RSS in the long-slit mode with a  1.5" slit and the PG0900 grating. This combination gives a spectral resolution of 5~\AA\ at a central wavelength of 5000~\AA\ (R $\sim$ 1000) and a wavelength coverage of 3320--6440~\AA.  We also used the grating PG1300 once to obtain the spectrum with a different instrumental set up.  The seeing conditions were typically $\sim$ 2" for the observing runs.  
 While a majority of the exposures had  an integration time of 1200 seconds, there are four exposures with an integration time of 900 seconds and one exposure with an integration time of 2100 seconds. With in a single observing run, we also obtained multiple spectra separated by few days/weeks to probe the short timescale variations.
Data reduction was performed using standard IRAF\footnote{IRAF is distributed by the National Optical Astronomy Observatories, which are operated by the Association of Universities for Research in Astronomy, Inc., under cooperative agreement with the National Science Foundation} scripts. The  preliminary  data  reduction  (gain  correction,  overscan bias subtraction, cross-talk correction and amplifier mosaicing) was done with  the SALT reduction pipeline. Subsequently, we  flat-fielded the data,  applied a wavelength solution using arc lamp spectra,   background subtracted the two dimensional spectra  and  extracted the one dimensional spectra around an aperture centered on the target. We also performed relative flux calibration on the data  using standard stars.  Absolute flux calibration was not performed due to the fixed nature of the SALT primary mirror and the strong variation in  effective aperture  with time  and source position. In addition, there are also slit losses due to seeing being of the order of or bigger than the slit width which introduce further uncertainty in the flux scale. The details of the SALT observations  are given in Table.~\ref{obs_tab}. As we did not find any detectable spectral variations between data obtained within a cycle (spanning $\sim$ 6 months ), we combined the individual exposures obtained within a cycle. 

We  obtained the continuum light curve measurements for SDSS J1333+0012 from the Catalina Real-Time Transient Survey \citep[CRTS,][]{drake09}.   CRTS operates with an unfiltered set up and the observed open magnitudes are transformed to V magnitudes using the  equation, ${\textrm{V} = \textrm{V}_{ins}+\textrm{a(V)}+\textrm{b(V)}*\textrm{(B-V)}}$, where V$_{ins}$ is the observed open magnitude. The zero-point a(V) and slope b(V) are obtained from three or more comparison stars in the same field. On a given night, The CRTS obtains four such observations taken 10 min apart.  Our spectroscopic observations have a good overlap with the CRTS observations.
\begin{figure}
 \centering
\includegraphics[width=1.0\linewidth,height=1.0\linewidth,angle=0]{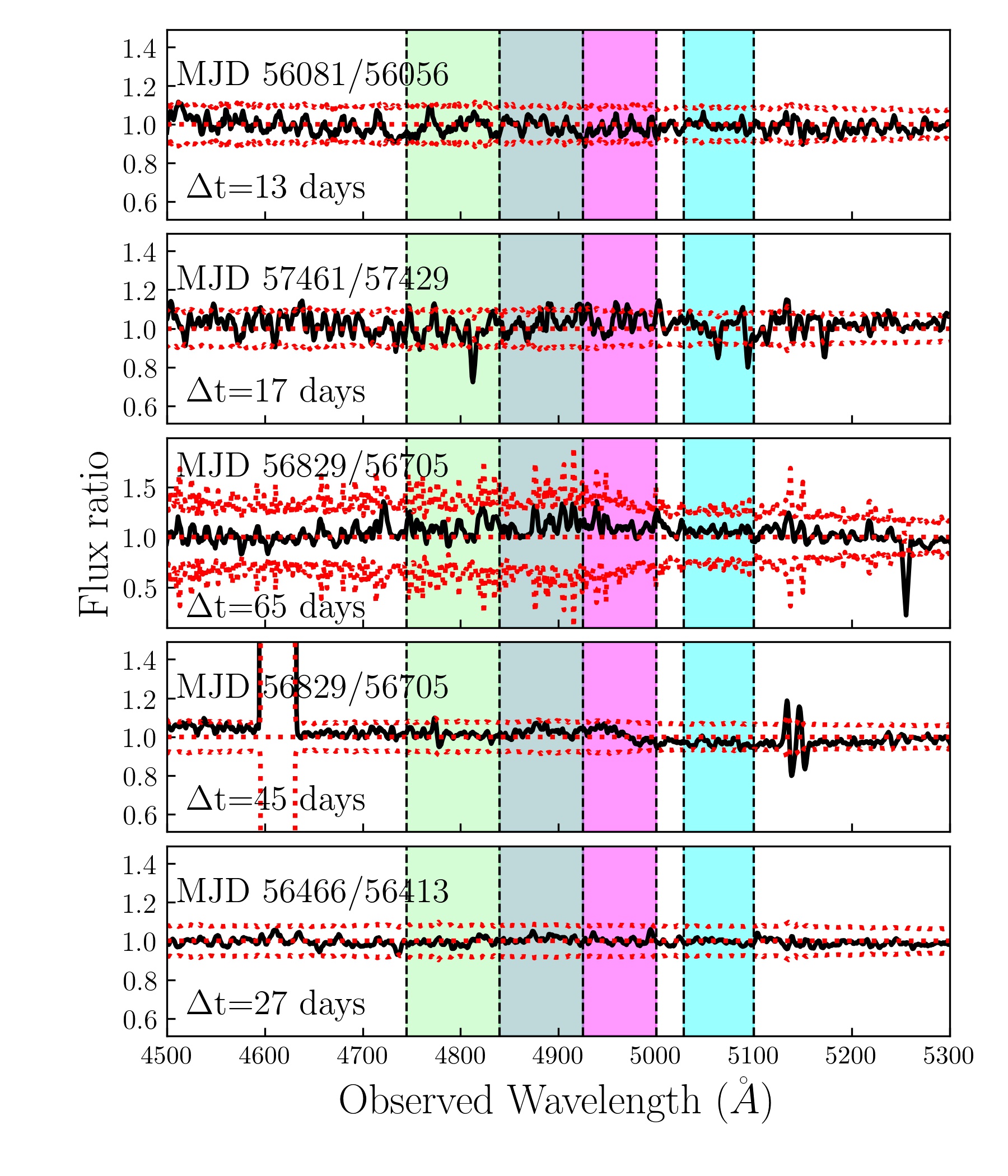}
\caption{ Short time-scale \mgii\ absorption line variations in the SALT spectra of J1333+0012. Each panel shows the ratio spectra for each year obtained by taking the ratio of two spectra which have the maximum time separation between the observations for a given year.  The associated 3$\sigma$ error on each ratio spectra are shown as red dotted lines. The MJDs of  the two observations and the time difference between the two in the QSO frame are marked in each panel. }
\label{ratio_shorttimescales_plot}
\end{figure}
\begin{figure*}
 \centering
\includegraphics[width=1.0\linewidth,height=1.\linewidth,angle=0]{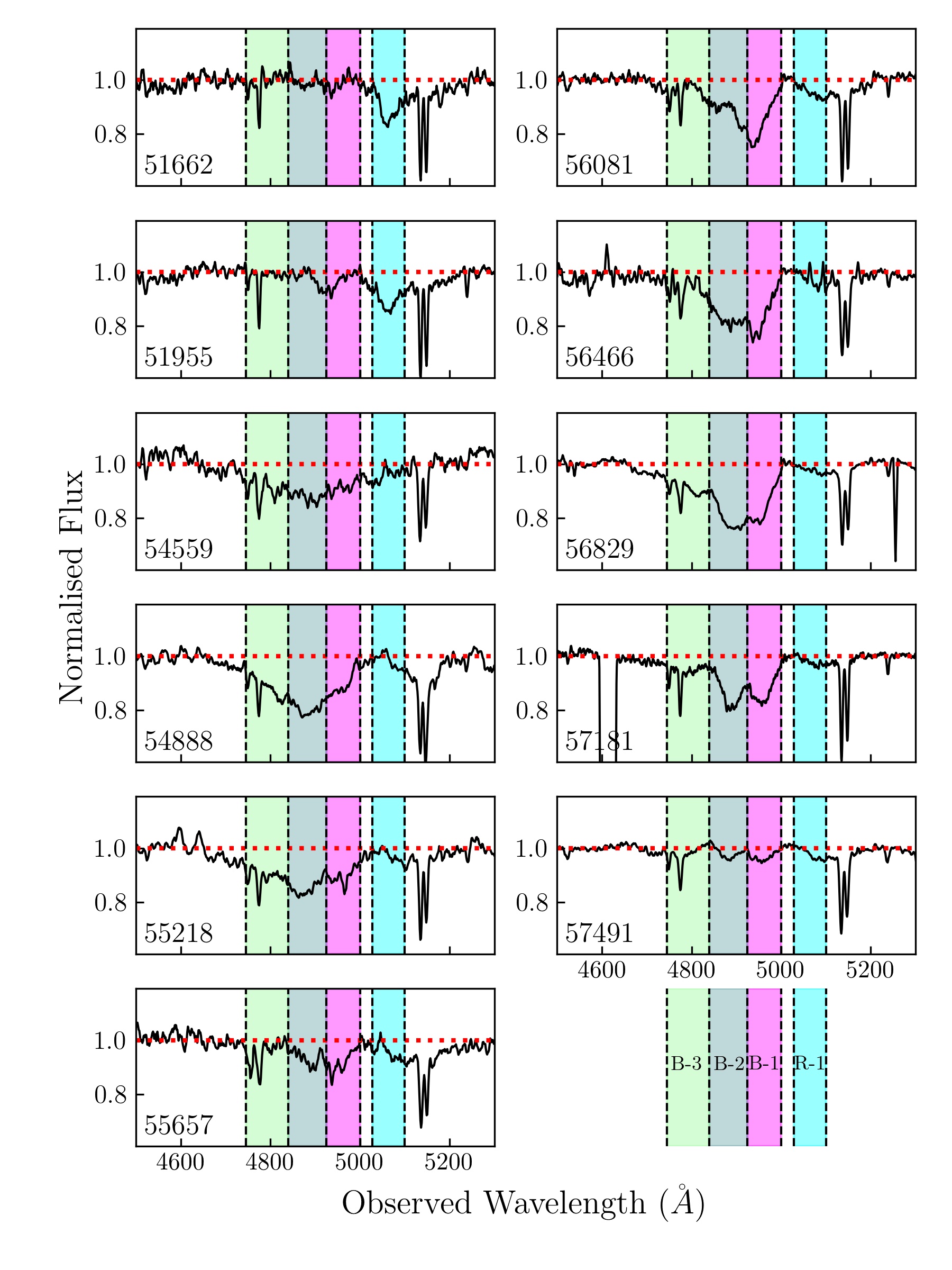}
\caption{ Comparison of continuum normalized spectra of all the epochs. The wavelength regions of the three B components and the R component are shaded in the figure by green, gray, blue and cyan. In each panel, the number in the lower left denotes the average MJD corresponding to the spectrum.
 }
\label{norm_plot}
\end{figure*}
 \begin{table*}
\caption{Equivalent width measurements for the three BAL components   }
\fontsize{7}{10}\selectfont
\begin{tabular}{|l|c|c c|c c|c c|c c|}
\hline
\hline
Instrument & MJD$^{a}$   &  \multicolumn{2}{c}{R Component}&\multicolumn{2}{c}{B-I Component}& \multicolumn{2}{c}{B-II Component}&\multicolumn{2}{c}{B-III Component}\\
\hline
           &       &   centroid& W$_R^{b}$       &   centroid&    W$_{B-I}^{c}$    &   centroid&W$_{B-II}^{d}$          &    centroid&W$_{B-III}^{e}$       \\
           &       &        ($\times$10$^3$\kms)           &     (\AA)          &    ($\times$10$^3$\kms)                  &       (\AA)       &      ($\times$10$^3$\kms)    &     (\AA)  &      ($\times$10$^3$\kms)  & (\AA)            \\                 
\hline
\hline

SDSS-2001 & 51662  &  $-$17.29$\pm$ 0.02     &   7.55$\pm$0.44  & $-$25.07$\pm$0.06    &     0.74$\pm$0.46 & $-$26.89$\pm$0.06    &     0.70$\pm$0.49 & $-$36.09$\pm$0.04   &      1.20$\pm$0.50   \\  
SDSS-2003 & 51955  &  $-$17.47$\pm$ 0.02     &   7.42$\pm$0.37  & $-$24.61$\pm$0.03    &     2.17$\pm$0.38 & $-$27.12$\pm$0.03    &     2.06$\pm$0.39 & $-$34.64$\pm$0.03   &      2.12$\pm$0.40   \\
IGO-2008  & 54559  &  $-$17.36$\pm$ 0.03     &   3.66$\pm$0.80  & $-$23.73$\pm$0.02    &     5.87$\pm$0.83 & $-$28.25$\pm$0.02    &     8.61$\pm$0.88 & $-$33.91$\pm$0.02   &      6.84$\pm$0.95   \\
IGO-2009  & 54888  &  $-$16.28$\pm$ 0.05     &   1.32$\pm$0.43  & $-$23.99$\pm$0.02    &     7.77$\pm$0.43 & $-$28.35$\pm$0.01    &    16.15$\pm$0.43 & $-$33.58$\pm$0.01   &     11.56$\pm$0.45   \\
IGO 2009  & 54916  &  $-$17.62$\pm$ 0.03     &   2.68$\pm$0.61  & $-$23.84$\pm$0.01    &    10.22$\pm$0.62 & $-$28.50$\pm$0.01    &    16.82$\pm$0.64 & $-$33.61$\pm$0.01   &     12.89$\pm$0.68   \\
IGO-2010  & 55218  &  $-$16.73$\pm$ 0.04     &   1.87$\pm$0.38  & $-$23.71$\pm$0.02    &     7.63$\pm$0.38 & $-$28.54$\pm$0.01    &    12.20$\pm$0.38 & $-$33.86$\pm$0.01   &     10.22$\pm$0.39   \\
IGO-2011  & 55657  &  $-$16.99$\pm$ 0.03     &   3.09$\pm$0.54  & $-$24.05$\pm$0.02    &     5.82$\pm$0.56 & $-$28.08$\pm$0.02    &     5.58$\pm$0.61 & $-$35.00$\pm$0.02   &      4.50$\pm$0.64   \\
SALT-2012 & 56081  &  $-$16.93$\pm$ 0.03     &   3.14$\pm$0.29  & $-$24.15$\pm$0.01    &    11.09$\pm$0.29 & $-$27.98$\pm$0.01    &    10.15$\pm$0.29 & $-$34.02$\pm$0.02   &      4.56$\pm$0.29   \\
SALT-2013 & 56466  &  $-$16.72$\pm$ 0.04     &   1.49$\pm$0.31  & $-$24.10$\pm$0.01    &    11.73$\pm$0.31 & $-$28.23$\pm$0.01    &    14.47$\pm$0.32 & $-$33.85$\pm$0.02   &      5.79$\pm$0.35   \\
SALT-2014 & 56829  &  $-$16.95$\pm$ 0.04     &   1.56$\pm$0.29  & $-$24.02$\pm$0.01    &    11.67$\pm$0.29 & $-$28.05$\pm$0.01    &    16.43$\pm$0.30 & $-$33.75$\pm$0.01   &      9.28$\pm$0.30	   \\
SALT-2015 & 57181  &  $-$16.70$\pm$ 0.03     &   1.40$\pm$0.34  & $-$23.80$\pm$0.01    &     9.54$\pm$0.34 & $-$28.02$\pm$0.01    &    11.08$\pm$0.35 & $-$34.30$\pm$0.01   &      6.06$\pm$0.36   \\
SALT-2016 & 57491  &  $-$16.36$\pm$ 0.04     &   1.24$\pm$0.30  & $-$23.82$\pm$0.03    &     4.95$\pm$0.30 & $-$27.75$\pm$0.04    &     4.56$\pm$0.31 & $-$35.41$\pm$0.03   &      3.83$\pm$0.31   \\

\hline
\hline
\end{tabular}
 \begin{flushleft}
{ $^a$ mid point of all the exposures taken in a year. \\
$^b$ computed between 5028--5100~\AA\ ; while no absorption line similar to that seen in the SDSS spectrum reappeared anytime during our observations, we do measure non-zero equivalent widths probably due to continuum fitting residuals or from a shallow absorption centered around 5150~\AA\ unrelated to the original red component.
  \\ $^c$ computed between 4925--5000~\AA\ ; $^d$ computed between 4840--4925~\AA\ \\
$^e$ computed between 4745--4840~\AA\ 
} 
\end{flushleft}
 \label{W_table}
\end{table*} 

\section{BAL Variability : Analysis \& Results }
 We first probed for short time-scale (13--65 days in the QSO frame) BAL variations in the SALT spectra of J1333+0012. Fig.~\ref{ratio_shorttimescales_plot} shows the short time-scale variations seen in the BAL components.  Each panel shows the ratio spectra for each year obtained by taking the ratio of two spectra which have the maximum time separation between the observations for that year. The associated 3$\sigma$ error on each ratio spectra are shown as red dotted lines. It is clear from this figure that there are no  appreciable absorption line variations (i.e., more than 3$\sigma$ level) on short time-scales. Hence, we combined all the individual spectra observed with the same  observational set up and obtained within one observing run i.e., spectra separated by a few days up to  3 months in the observed frame. We assign the average MJD as the MJD corresponding to the combined spectrum. This resulted in 10 high signal-to-noise spectra spanning  the period 2008 to 2016. These spectra together with the two archival spectra available from the SDSS form the basis of our long time-scale (155 days--16 years in quasar frame) BAL variability analysis presented in this paper.

 We then fitted  the continuum  to all the spectra. The fitting procedure involves masking the wavelength range of  bad pixels and BAL absorption (4644$\le \lambda \le$5237~\AA), and fitting the flux in the remaining pixels with a second order polynomial.  In the case of BAL quasars, continuum measurements are difficult as the spectra are dominated by broad absorption. The narrow absorption lines seen in QSO spectra are found to be stable over long timescales. J1333+0012 has two sets of narrow intervening absorption lines associated with two \mgii\ systems at \zabs = 0.8362 and 0.8984  on either sides of the BAL trough.  We overcome the difficulty in continuum fitting by iterating  the above continuum fitting procedure. The best continuum is chosen to be the one which produces minimum variations in the equivalent width of narrow intervening \mgii\ absorption lines. The standard deviation of the equivalent width values of narrow \mgii\ components  for each best fit continuum procedure (0.277~\AA) is taken to be the error associated with the continuum fitting procedure. We then normalized each epoch spectrum with the corresponding best fitted continuum model.

Fig.~\ref{norm_plot} shows the comparison of continuum normalized spectra of all the epochs. As explained in \cite{vivek12a}, we define two BAL components, namely "blue" (hereafter, B component) and "red" (hereafter, R component).  The velocity limits of the B and R components were defined as the velocity after which the normalized flux rises above 0.9.  The higher signal-to-noise ratio (SNR) SALT spectra allow us to further resolve the B component. The SALT spectrum obtained in the year 2015 showed the presence of three sub components. As our aim is to probe the BAL variations in different velocity sub-bins, we did not attempt to algorithmically define the velocity edges of the sub components. Rather, we  divided the B component into three different components namely B-I, B-II and  B-III based on our visual inspection of the 2015 SALT spectrum. The R component originally detected in the early SDSS spectra never reappeared in our follow up SALT observations.  The wavelength regions of the three B components and the R component are shown in the figure by green, gray, blue and cyan shaded regions. It is evident from Fig.~\ref{norm_plot} that all  three B components show  variations in the absorption strength between any consecutive year data. During our IGO monitoring campaign, the B component BAL absorption had a maximum strength in 2009 (i.e., MJD 54888) and nearly disappeared in 2011 (i.e., MJD 55657). We see the same trend of strengthening and fading for all the three components in our SALT monitoring campaign. The B components hit a maximum strength in 2014 (i.e., MJD 56829) and then nearly disappeared in our latest 2016 (i.e., MJD 57491) observations.   \citet{mcgraw17} reported a fractional equivalent width change higher than 1.5 for the disappearing BAL troughs in their sample. In this study, we note that BAL trough depths did not completely reach the continuum even in 2011 and 2016 when the trough depths were minimum.  The fractional equivalent width changes in 2011 and 2016 are  1.0 and 1.2 respectively. Hence, we only claim a near disappearance rather than a complete disappearance for the B components.  We notice that the B component variations happen over a time period of 3 years in the observed frame. This correspond to a time-scale of 1.56 years in the rest-frame of the quasar.
\begin{figure}
 \centering
\includegraphics[width=1.0\linewidth,height=1.0\linewidth]{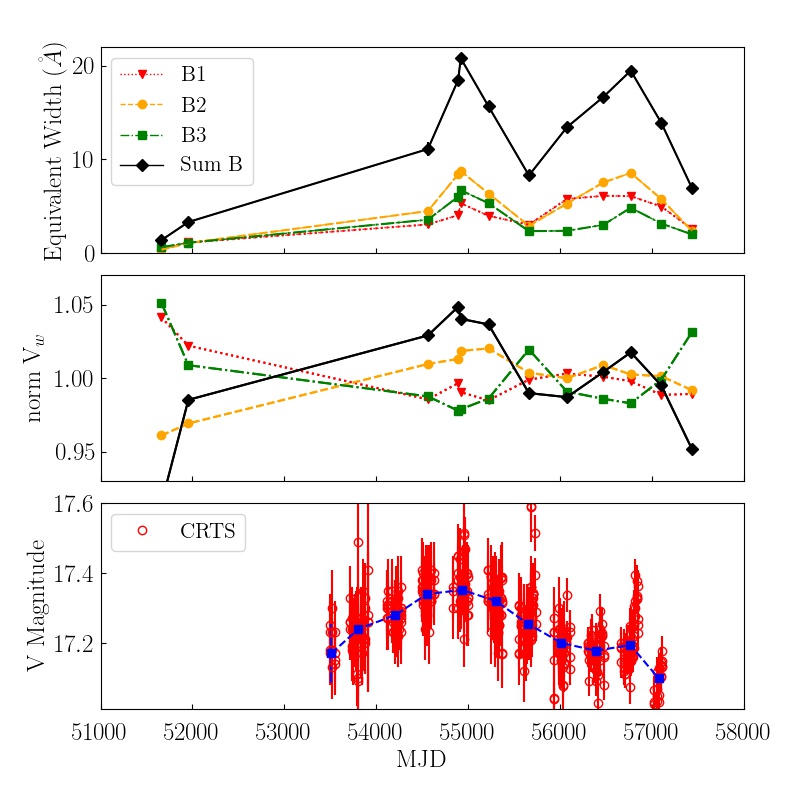}
	\caption{ The top panel shows the variation of \mgii\ equivalent widths with time for  different BAL components. The  middle panel shows the variation of normalized optical depth weighted velocity centroids, V$_{w}$, of different \mgii\ BAL components. The bottom panel shows the variation of CRTS V band magnitude over the same time.  The blue dashed line in the bottom panel corresponds to the median magnitude averaged over 100 days. 
 }
\label{W_plot}
\end{figure}
We computed  \mgii\ equivalent widths for all  four components in each  spectrum. The wavelength regions used for the measurement and the resulting equivalent widths  are listed in Table~\ref{W_table}. We do measure non-zero equivalent width while integrating the transmitted flux over the red region. This could be related to the residual from the continuum fitting or due to a shallow absorption centered around 5150~\AA\ unrelated to the red component seen in the SDSS spectrum. While we provide the measured equivalent width for this wavelength range, our focus in this paper is mainly on the blue component. The top panel of Fig.~\ref{W_plot} shows the variation of \mgii\ equivalent widths with MJD.  The total  as well as the individual  equivalent width measurements for the identified regions of the blue \mgii\ absorption line components clearly show the rapid  variation observed  within our observation campaign.    We also computed the optical depth weighted velocity centroids, V$_{w}$, of \mgii\ BAL absorption for all epochs. The middle panel shows the variation of optical depth weighted velocity centroid with MJD. We normalized the optical depth weighted velocity centroid of each absorption component by its mean velocity centroid(V($\tau$)) to compare the variation in different components. The optical depth weighted velocity centroids do not show large variations pointing to the velocity stability  of BAL troughs. In the bottom panel of Fig.~\ref{W_plot}, we show the CRTS  light curve. In the period over which the light curve is plotted, we see the quasar dimming  a bit before brightening. However the maximum change is $\sim$ 0.3 mag. The light curve does not show any significant double peak seen in the equivalent width plot.

\begin{figure}
 \centering
\includegraphics[width=1.0\linewidth,height=0.8\linewidth]{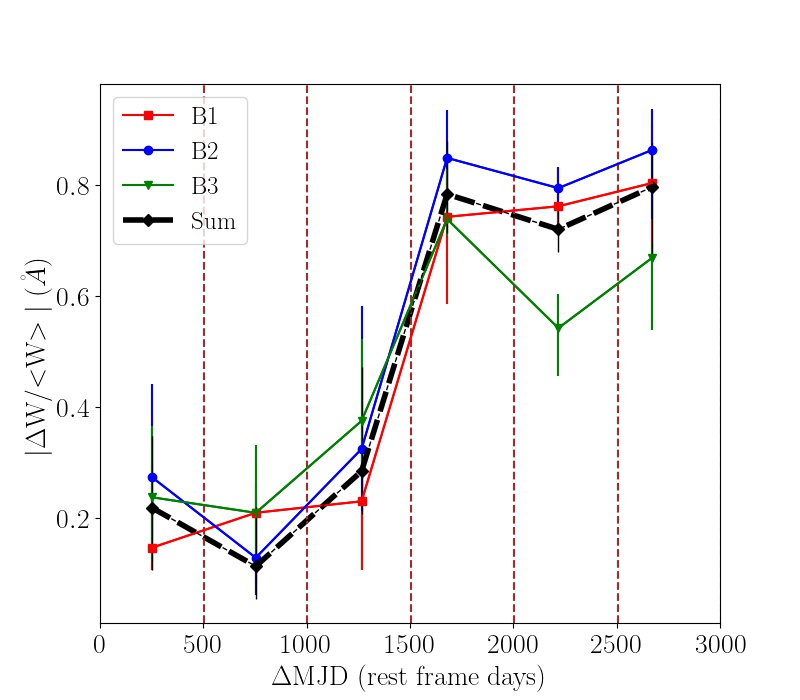}
\caption{ Median value of  absolute fractional equivalent width variation of the blue \mgii\ components binned in 500 days is plotted against the rest-frame time lags. The  upper and lower error bars correspond to the 75 and 25 percentiles. The dashed vertical lines mark the boundaries of the time-lag bins.
 }
\label{W_frac}
\end{figure}
\begin{figure*}
 \centering
\begin{tabular}{c}

\includegraphics[width=1.0\linewidth,height=0.6\linewidth]{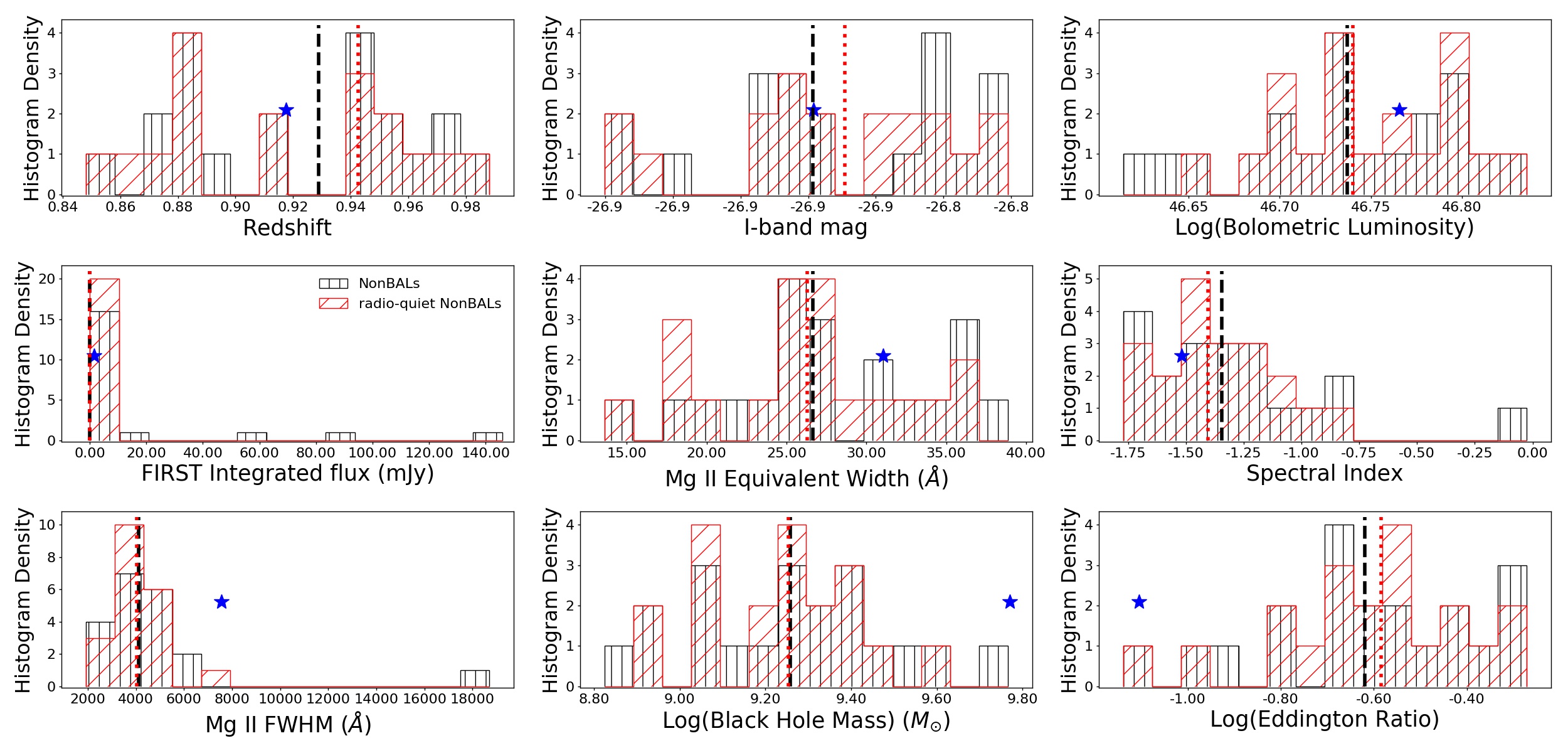}

\end{tabular}
\caption{The histogram distributions of the various QSO parameters for the two control sample described in section 3.1. The  vertical and slanted hatched histograms represent the nonBAL and radio-quiet nonBAL control sample respectively.  The dashed green and dotted red lines represent the median of the nonBAL and radio-quiet nonBAL distributions respectively. The blue star represents the parameter value for J1333+0012. }
\label{control_nonbal_plot}
\end{figure*}
\begin{figure}
 \centering
\includegraphics[width=1.0\linewidth,height=0.8\linewidth]{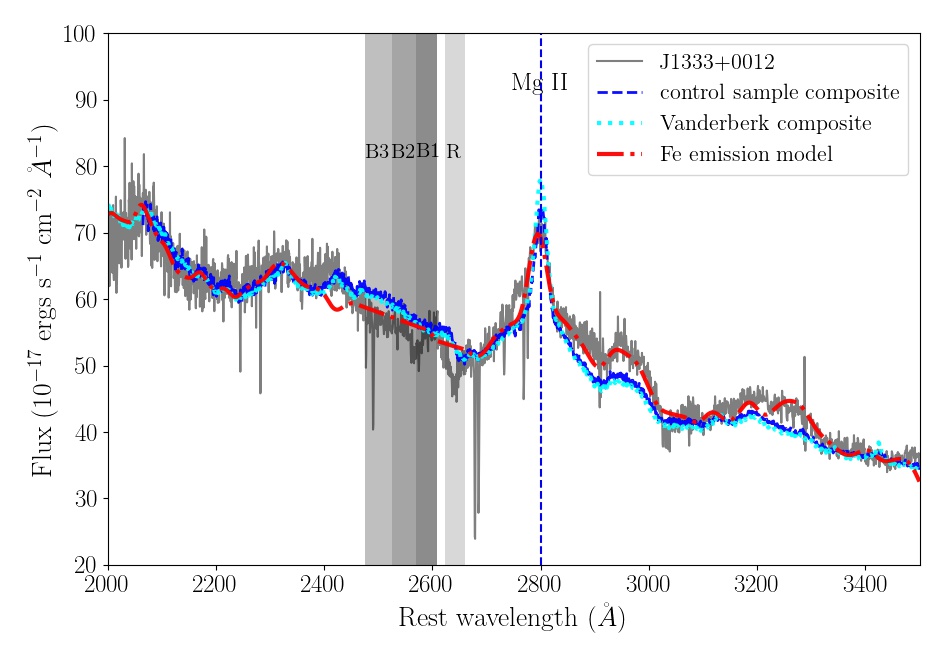}
\caption{ Comparison of J1333+0012 SDSS spectrum (solid gray) with the composite spectrum generated from the DR7  control sample (dashed/blue).  The \citet{vandenberk01} non-BAL composite is shown as dotted/cyan line. The dot dashed/red line shows the \citet{vestergard01} Fe emission model fitting together with a power law  and double Gaussian. 
 }
\label{composite_plot}
\end{figure}

Fig.~\ref{W_frac} shows the variation of absolute fractional change in equivalent width, defined as $|\Delta$ W$|$/$<$W$>$, as a function of rest-frame time-lag for the blue absorption components. Each data point in Fig.~\ref{W_frac} represents the median value of the  absolute fractional change in equivalent widths binned by 500 rest-frame days. The lower and upper error bars correspond to the 25 and 75 percentile of the  $|\Delta$ W$|$/$<$W$>$ distribution within the bin. The dashed vertical lines mark the boundaries of the bins. Clearly, there is a difference in the amplitude of BAL variation between short timescales ($<$ 1000 days) and long timescales ($>$ 1500 days).  Our spectra cover only the wavelength range of \feii\ lines associated with the broad \mgii\ absorption. Neither in the individual spectrum nor in the combined spectrum, we detect \feii\ lines associated with the broad \mgii\ absorption.
\subsection{Control sample of non-BALQSOs}
The peculiar variations of J1333+0012 motivated us to construct a control sample of non-BALQSOs and to compare the various observational properties of J1333+0012 with that of the control sample QSOs. A control sample of 20 non-BALQSOs was identified from the SDSS DR7 QSO properties catalogue \citep{shen11} using the k-nearest neighbor algorithm, which minimizes the distance in the redshift, absolute luminosity plane.  Radio-loud quasars are often considered as a distinct group within the BALQSO sample. J1333+0012 has an integrated flux of 2.43 mJy at 1.4Ghz in the FIRST catalog. In terms of radio loudness parameter \citep{kellerman89}, J1333+0012 has a value of 3.35 and would not qualify as a radio-loud quasar. Following the procedure for the previous control sample, we also constructed another control sample containing only radio-quiet quasars. 
 Fig.~\ref{control_nonbal_plot} shows the histogram distributions of the various QSO parameters for the objects in the two control samples. The dashed green  and dotted red lines represent the median of the nonBAL and radio-quiet nonBAL distributions respectively, and the blue star represents the parameter value for J1333+0012.  The different panels represents the histogram distributions of redshift, absolute i-band magnitude, bolometric luminosity, FIRST integrated radio flux, equivalent width of \mgii\ emission line,  the power-law spectral index measured near the \mgii\ emission line, FWHM of \mgii\ emission line, black hole mass estimated from \mgii\ emission line and the resulting Eddington ratio. Except for the parameters in the third row, namely FWHM, black hole mass and Eddington ratio, the parameter values for J1333+0012 are distributed around the median of the distribution for the two control samples. The distribution of parameters between the two control samples matches well with each other. We note that the FWHM of the \mgii\ emission line in J1333+0012 is slightly higher compared the FWHMs of \mgii\ emission line in control sample QSOs. Black hole mass and Eddington ratio are derived from the \mgii\ FWHM measurements.  A control sample comprising of \mgii\ BALs have larger dispersion in the redshift, and absolute i-band magnitude distributions as the fraction of LoBALs are much lower than HiBALs. With this caveat in mind, we note that the distribution of different parameters in the \mgii\ BAL control sample matches well with the non-BAL control sample.

We used this  control sample of 20  non-BALQSOs to generate a composite spectrum. To assemble the composite, we normalized  each individual spectrum   at 2500~\AA\ and also distorted the spectral indices to have the same mean value. We also generated another composite from a control sample of QSOs selected from BOSS DR12 catalogue \citep{paris14}.  We find that the  DR7 and DR12  composites are sufficiently similar to each other that we use the DR7 composite to compare with the J1333+0012 spectrum as we have access to other QSO properties from the \citep{shen11} catalogue.    Fig.~\ref{composite_plot} shows the comparison of J1333+0012 SDSS spectrum with the composite spectrum generated from the DR7  control sample QSOs (dashed/blue line).  The \citet{vandenberk01} composite is also shown as dotted/cyan line. Our composite spectrum generated from the DR7 control sample QSOs is similar to the \citet{vandenberk01} non-BAL composite. Both the DR7 control sample composite and the \citet{vandenberk01} non-BAL composite fail to fit the features red-ward of the \mgii\ emission line. We then fit the J1333+0012 spectrum using  three components: a power law component, a double Gaussian component to fit the \mgii\ emission line and a Fe emission template from  \citet{vestergard01}.  While fitting, we masked the wavelengths corresponding the the \mgii\ BAL components (shaded region).    The dot dashed/red line shows the \citet{vestergard01} Fe emission model fitting together with a power law  and a double Gaussian. The Fe emission model better fits spectral features red-ward of the \mgii\ emission line suggesting that J1333+0012 spectrum has significant contributions from iron emission.   As has been previously noted, the \mgii\ emission line of J1333+0012 appears to be slightly broader than the DR7 composite.

\section{Discussion \& Summary}
 The  main  result  of  our  monitoring  of  J1333+0012  is  that  the three blue components of \mgii\ absorption appear to vary  in equivalent widths in phase over an extended period of time. Unlike the blue components, the red component, originally  seen in the SDSS epochs, never reappeared  during our spectroscopic monitoring  campaign.  In \citet[][]{vivek12}, we explored the scenario where one lower velocity component (original red component) got accelerated to a higher velocity component \citep[for details about acceleration scenario, see section 4 of ][]{vivek12}. The reappearance of the blue components at the same velocities in the new SALT observations do not support the acceleration scenario.  

It may be possible that we are observing J1333+0012  during a  special time in its lifetime when it is trying to drive an outflow during its infancy. Evolutionary models for quasar outflows indeed have such a phase in the initial stages   when the quasar is trying to blow off the dust and gas coccoon \citep[for e.g.,][]{boroson92,becker00,urrutia09}.   Outflows facilitate the accretion onto the black hole by removing the angular momentum of the gas in the accretion disk. If  the angular momentum removed by the outflow is equal to the angular momentum transferred from the accretion, one can obtain a linear relation between the outflow rate and the accretion rate \citep{konigl00}.  Episodic outflow ejections are reported in the case of young stellar objects(YSOs) which represent the earliest stages of stellar evolutions \citep[for e.g.,][]{bell94,vorobyov05}.  Similarly, the appearing/reappearing outflows in J1333+0012 may be powered by episodic accretion events during the initial phases of quasar evolution. However,  the reappearance of the BAL trough at the same velocity is less probable in the case of outflows powered by episodic accretion events.  Hence, we do not favor the episodic outflow ejection model. 
 Here, we explore various other scenarios which can explain the absorption line variability noted here.

\subsection{Multiple streaming clouds across line of sight in a co-rotating wind:} In the physical scenario of magnetocentrifugal driven winds \citep{dekool95,proga03,everett05}, winds co-rotate with the disk, at least close to the disk.  Such a wind is made of dense clouds confined by the magnetic field, and therefore does not require shielding. Two of the predictions of this co-rotating wind scenario are high terminal velocities of the outflow and broader emission lines as compared to a non-rotating wind.   We note that J1333+0012 has a maximum ejection velocity of 32000~\kms while the average maximum ejection velocity in \citet{vivek14} is $\sim$ 5800~\kms \citep[see fig 11 of][]{vivek14}. Thus, the velocity of J1333+0012 is  on the higher side of the observed ejection velocities in BALQSOs.  From Fig.~\ref{control_nonbal_plot} and Fig.~\ref{composite_plot}, we also note that the \mgii\ emission line in J1333+0012 is broader than \mgii\ emission lines in the control sample QSOs. The FWHM of the \mgii\ emission line is 7552~\kms whereas  the non-BALQSO, radio-quiet non-BALQSO and LoBAL QSO control sample have a median value of 4000, 4100 and 4300~\kms.  

Assuming that BAL material is launched from a rotating disk, one would expect the disk wind to continue to rotate as it travels outward. We can estimate the dynamics of this rotating wind from the measured properties of the quasar.    We obtained the B band magnitude of this QSO using the measured u and g 
magnitudes \citep[see][]{Jester05}.  This B band magnitude together with the prescription 
of \citet{marconi04} results in the bolometric luminosity,  
L$_{bol}$ = 9.39 $\times$ 10$^{46}$~ergs s$^{-1}$. The corresponding 
mass accretion rate,$\dot{M_{in}}$ is,
\begin{equation}
\centering
 \dot{M_{in}} = \frac{L_{bol}}{\epsilon c^2} = 16.9 M_{\odot}/yr 
 \end{equation}
 \label{eqn4}
where $\epsilon$, the mass to energy conversion efficiency is taken as 0.1.  
 The corresponding blackhole mass assuming an Eddington accretion is 8$\times 10^8$ M$_\odot$.

 If the measured BAL radial velocity of 25000~\kms\ is assumed to be close to the actual three dimensional velocity vector of the BAL cloud, a 1$\times 10^9$ M$_\odot$ blackhole indicates a  Keplerian circular orbital period $\sim$ 2 years. We note that \citet{shen11} report the blackhole mass for J1333+0012 as 6$\times 10^9$ M$_\odot$ and this blackhole mass estimate results in an orbital period of $\sim$ 10 years. The maximum rest-frame timescale probed by our observations is 4.2 years. This would mean that the BAL cloud would have  completed a significant fraction of a full rotation during our observations and the cloud would have moved out of the line of sight in {1--2} years timescale. In our observations, we do find that the BAL troughs  nearly disappeared on a rest-frame timescale of 2 years. However, we do not favour the scenario where the same cloud reappearing again as the timescale between near disappearance and subsequent reappearence is within a rest-frame year. The  Keplerian circular orbit may be a too simple of an assumption given that the line of sight is very large. In reality, the cloud may be moving in an elongated elliptical orbit and the measured radial velocity may not be close to the true velocity of the BAL cloud. In an elongated elliptical orbit, if the cloud is moving in a direction close to the line of sight,  the radial velocity will dominate over the transverse velocity. \citet{vivek12} measured the transverse velocity of the BAL cloud to be 550~\kms\. The small value of transverse velocity  as compared to the radial velocity would imply that the BAL cloud is indeed moving close to the line of sight in an elongated orbit.

 When multiple clouds rotate with the disk, the distribution of clouds   in the line of sight changes with time.      In this scenario the strengthening and weakening of BAL absorption can be attributed to the  covering fraction changes of the passing BAL clouds. The minimum of BAL absorption strength may be the case when the line of sight does not intersect any BAL cloud. Subsequent reappearance and evolution of BAL absorption may be  attributed to the continued covering fraction changes of the  BAL clouds. Although the reappearance of the second cloud at the same velocity seems less likely, it can happen if the clouds are density structures embedded in a bigger outflow. 

\subsection{Photoionization driven BAL variability:} Coordinated variations over large velocities may suggest that BAL variability is caused by changes in the ionization state of the absorbers. The photoionization induced variability can be well studied either when two or more absorption components from the same ion are detected or when there are absorption from different ions with similar ionization potential. As there is only a single \mgii\ BAL component in our spectra, the only handle we have is to look for coordinated variations between continuum and absorption lines. However, previous studies on BAL variability have not detected any correlations between BAL absorption and continuum variations \citep{gibson08,filiz13,vivek14}. If the density of the absorbing cloud is assumed to be constant, change in the ionization state of the gas  can only happen with a change in the ionizing flux. Bottom panel of Fig.~\ref{W_plot} shows the CRTS V-band light curve obtained roughly during the same time of our spectroscopic observations.   The QSO continuum flux varied in concordance with the BAL variability during the first phase of BAL variability (i.e, between 2008 and 2011). When the BAL absorption got stronger, the QSO flux dimmed and vice versa.   In \citet{vivek12}, we speculated this possible connection between QSO dimming and strengthening of \mgii\ equivalent widths to some outflow ejection events in the accretion disc which cause reduction in the accretion efficiency. However, a similar concordant variability is not seen in the second phase of our monitoring (i.e., between 2012 and 2016).

\begin{figure}
 \centering
\includegraphics[width=1\linewidth,height=0.6\linewidth]{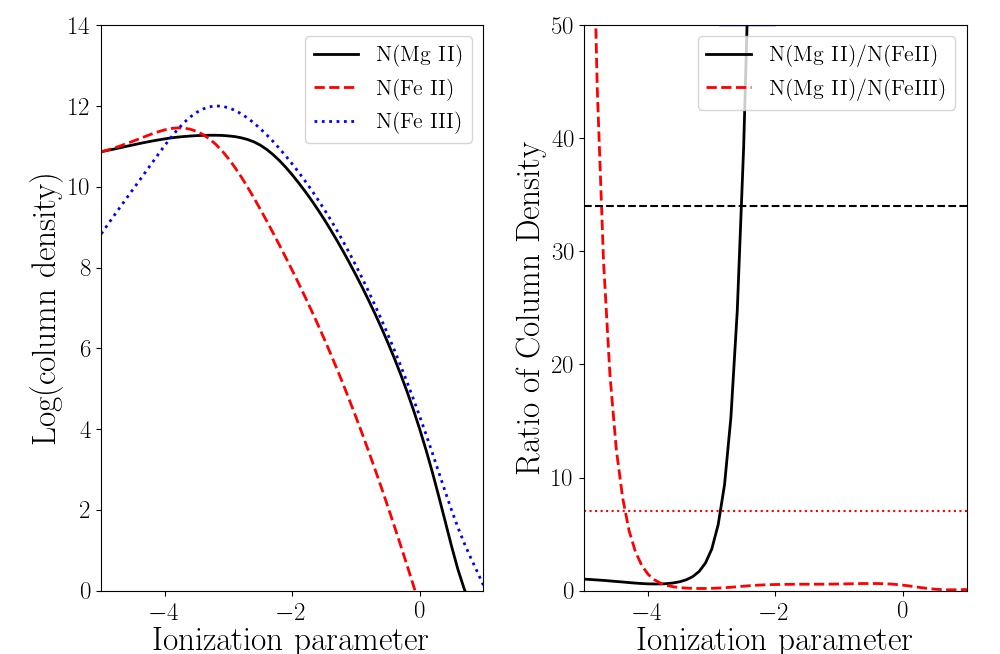}
\caption{ The left panel shows the column densities of \mgii\ (black/solid), ground state \feii\ (red,dashed) and ground state \feiii\ (blue/dotted) lines as a function of ionization parameter. The right panel shows the  ratio of column densities of ground state \feii\ (black/solid) and ground state \feiii (red/dashed) with respect to \mgii\ column density. The dashed black line corresponds to the measured ratio of equivalent widths of \mgii\ and \feii\  and dotted red line corresponds to the measured ratio of equivalent widths of \mgii\ and \feiii.}
\label{cloudy_plot}
\end{figure}

We computed the limit on \feii\ line equivalent width corresponding to the velocity of  the \mgii\ blue component and then used {\scriptsize CLOUDY} v17.00 to  model the photoionization properties of the absorbing cloud. {\scriptsize CLOUDY} is a spectral synthesis and plasma simulation code designed to simulate  astrophysical environments \citep{ferland17}.   We found the 3$\sigma$ limit on the  ground state \feii\ equivalent width to be 0.6~\AA. \feiii\ excited fine-structure line UV 49 is also covered in the SALT spectra. The 3$\sigma$ limit for UV 49 line is found to be 1.5~\AA.  The left panel in Fig.~\ref{cloudy_plot} shows the column densities of \mgii\ (black/solid), ground state \feii(red/dashed) and ground state \feiii\ (blue/dotted) lines as a function of ionization parameter. The right panel shows the  ratio of column densities of ground state \feii\ (black/solid) and ground state \feiii\ (red/dashed) with respect to \mgii\ column density. The black/dashed and red/dotted  horizontal lines correspond to the lower limit on equivalent width ratios,  W(\mgii)/W(\feii) and W(\mgii)/W(\feiii) measured from the spectrum having strongest \mgii\ absorption (obtained in the year 2014). In computing the above equivalent width ratios, we have used the total equivalent width of the \mgii\ blue component rather using than the equivalent widths of the sub-components. As none of the \mgii\ lines are saturated, it is reasonable to approximate that the column densities and equivalent widths are linearly connected. Unfortunately, we do not detect \feiii\ ground state transitions in any of our spectra. The current version of {\scriptsize CLOUDY} cannot handle \feiii\ excited fine structure lines. So, we use the column densities of \mgii\ and \feii\ for our calculations. The \mgii\ to \feii\ column density ratio   points to an ionization parameter range between $-$4 to $-$3. The right panel of Fig.~\ref{cloudy_plot} shows that a small increase in ionization parameter around $-$3 can results in a  large increase  in the W(\mgii)/W(\feii) ratio. This would mean that a large change in \mgii\ column density without  any \feii\ absorption can be achieved by a small change in the ionization parameter.

 As our interest is in the long term variations of the light curve, we employed a median filtering of the light curve with a window size of 200 days. The maximum and minimum magnitudes of the median filtered light curve is 17.09 mag and 17.35 mag.   The QSO has only varied by $\sim$0.26 mag in the V band light curve. 
This small change in magnitude alone cannot  explain the observed large change in the \mgii\ equivalent width \citep[see Fig.2 of ][]{hamann97}. 
 It is possible that changes in the ionizing UV continuum is much more than the observed changes in the V-band. In that scenario, photoionization changes in the BAL clouds due to the changes  in the shielding gas can explain the observed variabilities in J1333+0012. 

Radiation driven wind models have postulated the existence of 'hitchhiking/shielding gas'  which provides the shielding for the absorbing gas to prevent it from becoming over-ionized \citep{murray95}. The shielding gas is located between the continuum source and the outflowing gas. The ionizing flux impinging upon the outflow is the transmitted continuum through the shielding gas.  In this model,  variations in this shielding gas regulates the  amount of ionizing continuum that reaches the absorbing gas \citep[e.g.,][]{arav15}, but does not necessarily affect the lower energy UV continuum. Variations in the shielding gas can be achieved either by a physical re-arrangement of the disc or by the  co-rotation of shielding gas with the disc. \citet{sim10} did  multidimensional hydrodynamical simulation of X-ray spectra for AGN accretion disc outflows and found out that the X-ray radiation scattered and reprocessed in the flow has an important role in determining the ionization conditions in the wind.  
\begin{figure*}
 \centering
 \includegraphics[width=1.0\linewidth,height=0.30\linewidth]{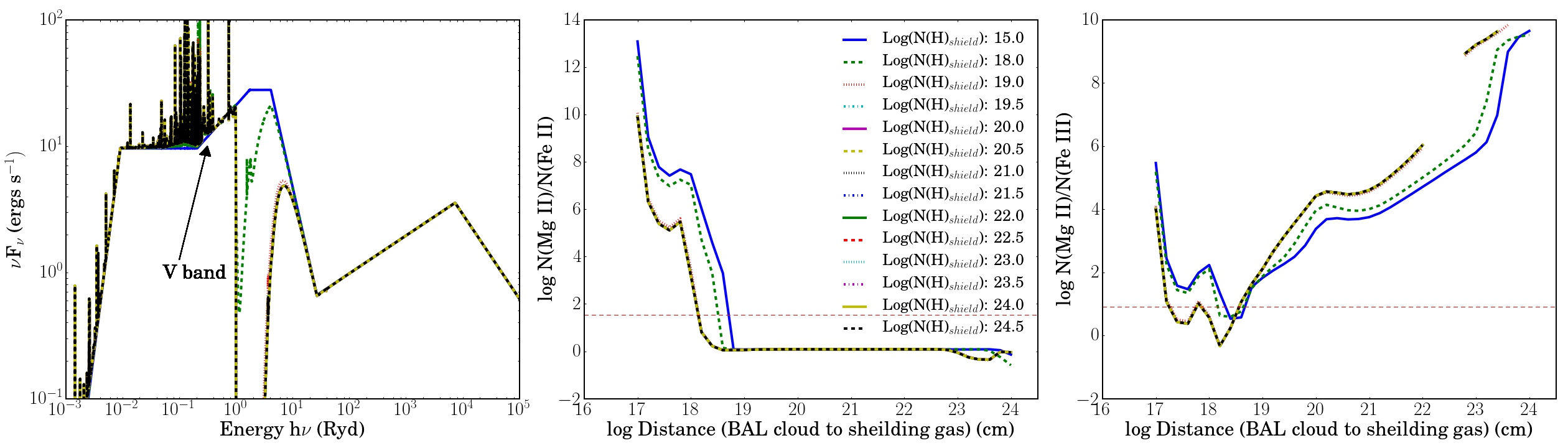}
	\caption{Left panel : Transmitted continua for various values of shielding gas column density.  The black arrow marks the location of the energy corresponding to the  V-band central wavelength. Middle panel: Variation of the ratio of \mgii\ to \feii\ column density as a function of distance between the BAL absorber and shielding gas for different values of hydrogen column density.  Right panel: Variation of the ratio of \mgii\ to \feiii\ column density as a function of distance between the BAL absorber and shielding gas for different values of hydrogen column density.  The red/dashed horizontal line in the middle and right panels corresponds to the observed upper limits on $N$(\mgii)/$N$(\feii) and $N$(\mgii)/$N$(\feiii).  }
\label{cloudy_plot2}
\end{figure*}

 We used {\scriptsize CLOUDY} simulations to test the hypothesis of a high column density ``shielding gas" regulating the ionization conditions of the absorbing cloud.
In this model, the shielding gas is located in between the continuum source and the BAL absorber.  Recent line driven  outflow simulations \citep{dyda18a,dyda18b} have predicted the existence of non-axisymmetric density features (clumps) at the base of the outflow. In the optically thick case, these clumps can  affect the outflow by decreasing the available ionizing flux and altering the ionization state of the outflowing gas. The density of these clumps also differs by a factor of $\sim$ 3 from the azimuthal average. These clumps at the base of the outflow can be thought of playing the role of shielding gas. We  ran two sets  of photoionization simulations  to test this hypothesis. In the first set of runs (hereafter, {\scriptsize CLOUDY-I} run) we used {\scriptsize CLOUDY} to obtain the continuum transmitted through the shielding gas. In the second set of runs (hereafter, {\scriptsize CLOUDY-II} run), we used the transmitted continuum from {\scriptsize CLOUDY-I} to study the photoionization conditions of the BAL cloud. In each of the {\scriptsize CLOUDY-I} run, we kept the flux of hydrogen-ionizing photons (log($\phi$(H)) = 12~cm$^{-2}$ s$^{-1}$) impinging the shielding gas cloud to be same and varied the total hydrogen column density of the shielding gas. This is equivalent to changing the shielding gas depth. We saved the transmitted continuum corresponding to the different values of the shielding gas column densities.  We normalized each transmitted continua to the same monochromatic luminosity of 46.5~ergs s$^{-1}$ at 0.1824~Ryd. This is to ensure that all the transmitted continua have the same power at a wavelength (beyond the main hydrogen absorption edge at  1~Ryd)  that was not absorbed in the first set of {\scriptsize CLOUDY} runs.  In {\scriptsize CLOUDY-II} runs, we used these normalized transmitted continua to determine the \mgii\ column density of the BAL absorber at different distances from the shielding gas. For both {\scriptsize CLOUDY-I} and {\scriptsize CLOUDY-II} runs, we assumed a hydrogen density of ${\textrm{n}_H = \textrm{10}^{10} \textrm{cm s}^{-3}}$ which is similar to the hydrogen density in the broad line region. The left panel of Fig.~\ref{cloudy_plot2} shows the different transmitted continua for various values of shielding gas column densities. When the shielding gas column density increases above 10$^{19}$~cm$^{-2}$, significant amount of hydrogen ionizing photons are absorbed by the shielding gas. The middle and right panel of Fig.~\ref{cloudy_plot2} show the ratio of column densities of \mgii\ to ground state \feii\ and \feiii\ respectively as a function of the distance of the BAL absorber from the shielding gas. The red/dashed horizontal line in the middle and right panels correspond to the measured upper limits on $\frac{N(\textrm\mgii)}{N(\textrm\feii)}$ and $\frac{N(\textrm \mgii)}{N(\textrm\feiii)}$.  It is clear from Fig.~\ref{cloudy_plot2} that the column density ratios of the BAL absorbing gas are sensitive to the column density of the shielding gas.  We also ran these simulations for higher values of $\phi(H)$ when the number of hydrogen-ionizing photons are much higher than the hydrogen density of the shielding gas. In all our simulations, we find that when the  BAL absorber distance is within a parsec from the shielding gas, the \mgii\ column densities can change by several orders of magnitude for a narrow range of shielding gas  hydrogen column density without producing  significant \feii\ and \feiii\ absorption.  A factor of 15 change in equivalent width in the B component on SDSS J1333+0012 can be explained by a change of 10$^{18}$ to 10$^{19}$~cm s$^{-2}$ in the column density of the shielding gas. The {\scriptsize CLOUDY} simulations reveal no significant change in the V-band flux due to this change in the shielding gas.  

The changes in the shielding gas hydrogen column density can be achieved through Keplerian rotation. The BAL absorbers are typically thought to have a launching radii of $\sim$ 1000 R$_s$. The Keplerian orbital period of a cloud at this radius is 17 years. As the location of shielding gas is in between the BAL absorber and the continuum source, the orbital period for the shielding gas will be even lesser. Thus, the shielding gas can have significant movement within our observed variability timescale of 1--2 years.  

We do not see any appreciable changes in the strength of the \mgii\ emission line between our observations. This would mean that the overall covering factor of the shielding gas is small. In Fig.~\ref{cloudy_plot2}, the energy corresponding to the redshifted V-band central wavelength is marked by the arrow. It is clear that the V-band continuum is not sensitive to the variations in the shielding gas. 

While shielding gas scenario is a viable option for the variability of  the blue absorption components, the same model will not explain the complete disappearance of the red component that did not reappear during our observations. This could either mean that the red and blue components are not co-located along our line of sight and have widely different physical conditions or more than one scenario is involved in the observed line variability. Further observations will shed more light on these issues.
 

\section*{Conclusion}
 In this paper, we report two cycles of  appearance and near disappearance in the \mgii\ broad absorption line outflow in SDSS J133356.02+001229.1. The blue component which appeared in 2001, is observed to first increase in absorption line strength in 2008. Reaching a maximum strength in 2009, it continued to decrease in strength and almost vanished in 2011. Furthermore, the absorption strength again started increasing in 2012, reached a maximum in 2014 and almost diminished in 2016.

Using CRTS light curves, we find that the quasar has not shown strong photometric variability that is correlated with the absorption line variability as expected in a simple photoionization scenario. However, these observations do not rule out a much larger variations in the ionizing continuum in the UV range. J1333+0012 has similar properties as that of a control sample of non-BALQSOs except for the FWHM of the \mgii\ emission line. The \mgii\ emission line FWHM of  J1333+0012 is slightly higher as compared to the control sample.
    
Using photoionization simulations, we argue that the observed variations of the blue components can be explained by the variable photoionization conditions of the outflow regulated by the  'shielding gas' located at the base of the outflow. The V-band continuum is not sensitive to the changes introduced by the shielding gas. Photometric monitoring in the UV will allow us to test this scenario as the UV flux variations are expected to be larger than optical variations. No variation in the \mgii\ emission line strength will be consistent with this scenario if the covering factor of the shielding gas is not very large. However, the variable shielding gas scenario cannot explain the  disappearance of the red component that has never reappeared during our monitoring period.  The observed absorption line variability  can also be    explained  by  multiple  streaming  gas  moving across our line of sight. But, the reappearance of the second cloud at the same velocity seems less probable. It is more likely that the actual scenario may be a combination of variable shielding gas and multiple streaming gas.  Continued monitoring of this source will be helpful to discern the actual nature of the outflow in J1333+0012.

\section*{acknowledgements}
The work of MV and KD was supported in part by the U.S. Department of Energy, Office of Science, Office of High Energy Physics, under Award Number DE-SC0009959. The support and resources from the Center for High Performance Computing at the University of Utah are gratefully acknowledged. { We thank the anonymous referee for a number of comments that helped us improve the paper.}
{

 
\def\aj{AJ}%
\def\actaa{Acta Astron.}%
\def\araa{ARA\&A}%
\def\apj{ApJ}%
\def\apjl{ApJ}%
\def\apjs{ApJS}%
\def\ao{Appl.~Opt.}%
\def\apss{Ap\&SS}%
\def\aap{A\&A}%
\def\aapr{A\&A~Rev.}%
\def\aaps{A\&AS}%
\def\azh{AZh}%
\def\baas{BAAS}%
\def\bac{Bull. astr. Inst. Czechosl.}%
\def\caa{Chinese Astron. Astrophys.}%
\def\cjaa{Chinese J. Astron. Astrophys.}%
\def\icarus{Icarus}%
\def\jcap{J. Cosmology Astropart. Phys.}%
\def\jrasc{JRASC}%
\def\mnras{MNRAS}%
\def\memras{MmRAS}%
\def\na{New A}%
\def\nar{New A Rev.}%
\def\pasa{PASA}%
\def\pra{Phys.~Rev.~A}%
\def\prb{Phys.~Rev.~B}%
\def\prc{Phys.~Rev.~C}%
\def\prd{Phys.~Rev.~D}%
\def\pre{Phys.~Rev.~E}%
\def\prl{Phys.~Rev.~Lett.}%
\def\pasp{PASP}%
\def\pasj{PASJ}%
\def\qjras{QJRAS}
\def\rmxaa{Rev. Mexicana Astron. Astrofis.}%
\def\skytel{S\&T}%
\def\solphys{Sol.~Phys.}%
\def\sovast{Soviet~Ast.}%
\def\ssr{Space~Sci.~Rev.}%
\def\zap{ZAp}%
\def\nat{Nature}%
\def\iaucirc{IAU~Circ.}%
\def\aplett{Astrophys.~Lett.}%
\def\apspr{Astrophys.~Space~Phys.~Res.}%
\def\bain{Bull.~Astron.~Inst.~Netherlands}%
\def\fcp{Fund.~Cosmic~Phys.}%
\def\gca{Geochim.~Cosmochim.~Acta}%
\def\grl{Geophys.~Res.~Lett.}%
\def\jcp{J.~Chem.~Phys.}%
\def\jgr{J.~Geophys.~Res.}%
\def\jqsrt{J.~Quant.~Spec.~Radiat.~Transf.}%
\def\memsai{Mem.~Soc.~Astron.~Italiana}%
\def\nphysa{Nucl.~Phys.~A}%
\def\physrep{Phys.~Rep.}%
\def\physscr{Phys.~Scr}%
\def\planss{Planet.~Space~Sci.}%
\def\procspie{Proc.~SPIE}%
\let\astap=\aap
\let\apjlett=\apjl
\let\apjsupp=\apjs
\let\applopt=\ao
\bibliographystyle{mn}
\bibliography{ref_rapid}
%

\end{document}